\documentstyle[12pt]{article}

\catcode`\@=11
\long\def\@makefntext#1{ 
\protect\noindent \hbox to 3.2pt {\hskip-.9pt
$^{{\ninerm\@thefnmark}}$\hfil}#1\hfill} 

\def\thefootnote{\fnsymbol{footnote}}
 \def\@makefnmark{\hbox to 0pt{$^{\@thefnmark}$\hss}}  

\def\ps@myheadings{\let\@mkboth\@gobbletwo
\def\@oddhead{\hbox{} 
\rightmark\hfil\ninerm\thepage}
\def\@oddfoot{}\def\@evenhead{\ninerm\thepage\hfil 
\leftmark\hbox{}}\def\@evenfoot{}
\def\sectionmark##1{}\def\subsectionmark##1{}}

\begin{document}

\newcommand{\symbolfootnote}{\renewcommand{\thefootnote}
	{\fnsymbol{footnote}}}
\renewcommand{\thefootnote}{\fnsymbol{footnote}}
\newcommand{\alphfootnote}
	{\setcounter{footnote}{0}
	 \renewcommand{\thefootnote}{\sevenrm\alph{footnote}}}

\newcounter{sectionc}\newcounter{subsectionc}\newcounter{subsubsectionc}
\renewcommand{\section}[1] {\vspace{0.6cm}\addtocounter{sectionc}{1}
\setcounter{subsectionc}{0}\setcounter{subsubsectionc}{0}\noindent
	{\bf\thesectionc. #1}\par\vspace{0.4cm}}
\renewcommand{\subsection}[1] {\vspace{0.6cm}\addtocounter{subsectionc}{1}
	\setcounter{subsubsectionc}{0}\noindent
	{\it\thesectionc.\thesubsectionc. #1}\par\vspace{0.4cm}}
\renewcommand{\subsubsection}[1] {\vspace{0.6cm}
\addtocounter{subsubsectionc}{1}
	\noindent {\rm\thesectionc.\thesubsectionc.\thesubsubsectionc.
	#1}\par\vspace{0.4cm}}
\newcommand{\nonumsection}[1] {\vspace{0.6cm}\noindent{\bf #1}
	\par\vspace{0.4cm}}

\newcounter{appendixc}
\newcounter{subappendixc}[appendixc]
\newcounter{subsubappendixc}[subappendixc]
\renewcommand{\thesubappendixc}{\Alph{appendixc}.\arabic{subappendixc}}
\renewcommand{\thesubsubappendixc}
	{\Alph{appendixc}.\arabic{subappendixc}.\arabic{subsubappendixc}}

\renewcommand{\appendix}[1] {\vspace{0.6cm}
        \refstepcounter{appendixc}
        \setcounter{figure}{0}
        \setcounter{table}{0}
        \setcounter{equation}{0}
        \renewcommand{\thefigure}{\Alph{appendixc}.\arabic{figure}}
        \renewcommand{\thetable}{\Alph{appendixc}.\arabic{table}}
        \renewcommand{\theappendixc}{\Alph{appendixc}}
        \renewcommand{\theequation}{\Alph{appendixc}.\arabic{equation}}
        \noindent{\bf Appendix \theappendixc #1}\par\vspace{0.4cm}}
\newcommand{\subappendix}[1] {\vspace{0.6cm}
        \refstepcounter{subappendixc}
        \noindent{\bf Appendix \thesubappendixc. #1}\par\vspace{0.4cm}}
\newcommand{\subsubappendix}[1] {\vspace{0.6cm}
        \refstepcounter{subsubappendixc}
        \noindent{\it Appendix \thesubsubappendixc. #1}
	\par\vspace{0.4cm}}

\def\abstracts#1{{
	\centering{\begin{minipage}{30pc}\tenrm\baselineskip=12pt\noindent
	\centerline{\tenrm ABSTRACT}\vspace{0.3cm}
	\parindent=0pt #1
	\end{minipage} }\par}}

\newcommand{\bibit}{\it}
\newcommand{\bibbf}{\bf}
\renewenvironment{thebibliography}[1]
	{\begin{list}{\arabic{enumi}.}
	{\usecounter{enumi}\setlength{\parsep}{0pt}
\setlength{\leftmargin 1.25cm}{\rightmargin 0pt}
	 \setlength{\itemsep}{0pt} \settowidth
	{\labelwidth}{#1.}\sloppy}}{\end{list}}

\topsep=0in\parsep=0in\itemsep=0in
\parindent=1.5pc

\newcounter{itemlistc}
\newcounter{romanlistc}
\newcounter{alphlistc}
\newcounter{arabiclistc}
\newenvironment{itemlist}
    	{\setcounter{itemlistc}{0}
	 \begin{list}{$\bullet$}
	{\usecounter{itemlistc}
	 \setlength{\parsep}{0pt}
	 \setlength{\itemsep}{0pt}}}{\end{list}}

\newenvironment{romanlist}
	{\setcounter{romanlistc}{0}
	 \begin{list}{$($\roman{romanlistc}$)$}
	{\usecounter{romanlistc}
	 \setlength{\parsep}{0pt}
	 \setlength{\itemsep}{0pt}}}{\end{list}}

\newenvironment{alphlist}
	{\setcounter{alphlistc}{0}
	 \begin{list}{$($\alph{alphlistc}$)$}
	{\usecounter{alphlistc}
	 \setlength{\parsep}{0pt}
	 \setlength{\itemsep}{0pt}}}{\end{list}}

\newenvironment{arabiclist}
	{\setcounter{arabiclistc}{0}
	 \begin{list}{\arabic{arabiclistc}}
	{\usecounter{arabiclistc}
	 \setlength{\parsep}{0pt}
	 \setlength{\itemsep}{0pt}}}{\end{list}}

\newcommand{\fcaption}[1]{
        \refstepcounter{figure}
        \setbox\@tempboxa = \hbox{\tenrm Fig.~\thefigure. #1}
        \ifdim \wd\@tempboxa > 6in
           {\begin{center}
        \parbox{6in}{\tenrm\baselineskip=12pt Fig.~\thefigure. #1 }
            \end{center}}
        \else
             {\begin{center}
             {\tenrm Fig.~\thefigure. #1}
              \end{center}}
        \fi}

\newcommand{\tcaption}[1]{
        \refstepcounter{table}
        \setbox\@tempboxa = \hbox{\tenrm Table~\thetable. #1}
        \ifdim \wd\@tempboxa > 6in
           {\begin{center}
        \parbox{6in}{\tenrm\baselineskip=12pt Table~\thetable. #1 }
            \end{center}}
        \else
             {\begin{center}
             {\tenrm Table~\thetable. #1}
              \end{center}}
        \fi}

\def\@citex[#1]#2{\if@filesw\immediate\write\@auxout
	{\string\citation{#2}}\fi
\def\@citea{}\@cite{\@for\@citeb:=#2\do
	{\@citea\def\@citea{,}\@ifundefined
	{b@\@citeb}{{\bf ?}\@warning
	{Citation `\@citeb' on page \thepage \space undefined}}
	{\csname b@\@citeb\endcsname}}}{#1}}

\newif\if@cghi
\def\cite{\@cghitrue\@ifnextchar [{\@tempswatrue
	\@citex}{\@tempswafalse\@citex[]}}
\def\citelow{\@cghifalse\@ifnextchar [{\@tempswatrue
	\@citex}{\@tempswafalse\@citex[]}}
\def\@cite#1#2{{$\null^{#1}$\if@tempswa\typeout
	{IJCGA warning: optional citation argument
	ignored: `#2'} \fi}}
\newcommand{\citeup}{\cite}

\def\fnm#1{$^{\mbox{\scriptsize #1}}$}
\def\fnt#1#2{\footnotetext{\kern-.3em
	{$^{\mbox{\sevenrm #1}}$}{#2}}}

\font\twelvebf=cmbx10 scaled\magstep 1
\font\twelverm=cmr10 scaled\magstep 1
\font\twelveit=cmti10 scaled\magstep 1
\font\elevenbfit=cmbxti10 scaled\magstephalf
\font\elevenbf=cmbx10 scaled\magstephalf
\font\elevenrm=cmr10 scaled\magstephalf
\font\elevenit=cmti10 scaled\magstephalf
\font\bfit=cmbxti10
\font\tenbf=cmbx10
\font\tenrm=cmr10
\font\tenit=cmti10
\font\ninebf=cmbx9
\font\ninerm=cmr9
\font\nineit=cmti9
\font\eightbf=cmbx8
\font\eightrm=cmr8
\font\eightit=cmti8

       {\normalsize \hfill
       \begin{tabbing}
       \`\begin{tabular}{l}
         SU-ITP-95-17  \\
         HUB-EP-95/22  \\
         hep--th/9510080 \\
         October 3, 1995 \\
         \end{tabular}
       \end{tabbing} }\vspace{8mm}
\thispagestyle{empty}

\centerline{\twelvebf String duality and massless string states}
\vspace{0.8cm}
\centerline{\tenrm Klaus Behrndt\footnote{Permanent address:
Humboldt--Universit\"at,
    Institut f\"ur Physik, 10115 Berlin, Germany \newline
    \hspace*{0in} E-mail: behrndt@qft2.physik.hu-berlin.de, Work
     supported by the DFG and a grant of the DAAD}}
\centerline{\tenit  Physics Department, Stanford University, Stanford   CA
94305-4060, USA}
\baselineskip=12pt
\vspace{0.9cm}

\renewcommand{\arraystretch}{2.0}
\renewcommand{\thefootnote}{\alph{footnote}}
\newcommand{\be}[3]{\begin{equation}  \label{#1#2#3}}
\newcommand{\ee}{\end{equation}}
\newcommand{\ba}{\begin{array}}
\newcommand{\ea}{\end{array}}
\newcommand{\vsf}{\vspace{5mm}}
\newcommand{\NP}[3]{{\em Nucl. Phys.}{ \bf B#1#2#3}}
\newcommand{\PRD}[2]{{\em Phys. Rev.}{ \bf D#1#2}}
\newcommand{\MPLA}[1]{{\em Mod. Phys. Lett.}{ \bf A#1}}
\newcommand{\PL}[3]{{\em Phys. Lett.}{ \bf B#1#2#3}}
\newcommand{\marpar}{\marginpar[!!!]{!!!}}
\newcommand{\lab}[2]{\label{#1#2}   (#1#2) \hfill }
\vspace{10mm}

\abstracts{
We are discussing the $S$ \& $T$ duality for special class
of heterotic string configurations. This class of solutions
includes various types of black hole solutions and Taub-NUT
geometries. It allows a self-dual point for both dualities which
corresponds to massless configurations. As string state
this point corresponds to $N_R=1/2$ and $N_L=0$. The string/string
duality is shortly discussed.
}
\vspace{10mm}

\vfill
\baselineskip=14pt

\begin{center}
Talk presented at the \\
{\em Workshop on strings, gravity and related topics} \\
Trieste, Italy, June 29 -- 30, 1995
\end{center}

\vfill

\newpage

\twelverm   
\baselineskip=14pt

Duality symmetries have played an important role for a better
understanding of different string backgrounds. They are given by field
redefinition and usually relates different geometries or even
different topologies what is unknown from point particle
physics. Mainly we have two types, one which are valid off-shell (as
symmetry of the low energy action), e.g.\ $T$ duality, or which are
valid on-shell only (as symmetry of the equation of motion), e.g.\ $S$
duality. Both dualities relate one heterotic background to another
one. But there are also duality transformations of completely
different kind. One example is the string/string duality in $D=6$
which transforms a heterotic background to type IIA background, and
hence relates different types of string theories to each other. For a
nice review of this topic see Ref.\cite{du}. To be correct one has to
admit that as well the string/string duality as the S-duality are
still conjectures, although the evidence for the latter one is
overwhelming.

The usual starting point in discussing dualities is the 10-dimensional
effective action
\be010
\ba{r}
S_{10} = \int d^{10} X \sqrt{\hat{G}} e^{- 2 \hat{\phi}} \left[ \; R +
        4 (\partial \hat{\phi})^2 - \frac{1}{12} \hat{H}^2
	          \; \right] +
    \cal{O}(\alpha') + \mbox{(higher genus terms) } \ .
\ea
\ee
Assuming that the fields do not depend on 6 coordinates (internal)
we can reduce this action down to 4 dimensions (for details see
e.g.\ Ref.\cite{se1})
\be020
\ba{r}
S_{4} = \int d^{4} x \sqrt{G} e^{- 2 \phi} \left[ \; R +
        4 (\partial \phi)^2 - \frac{1}{12} H^2 + \mbox{KK-gauge fields} +
	\mbox{KK-scalars} \right] +\\
      +  \cal{O}(\alpha') + \mbox{(higher genus terms)}     \ .
\ea
\ee
The orign of the T-duality is a space time symmetry of the 10d theory,
namely the existence of a Killing vector. Let us assume that $u$ is the
symmetry direction and $I,J$ are the other coordinates, then the T-duality
can be written as \cite{bu}
\be030
\ba{l}
G_{uu} \rightarrow 1/G_{uu} \quad , \quad G_{uI} \rightarrow
B_{uI}/G_{uu} \quad , \quad \quad B_{uI} \rightarrow G_{uI}/G_{uu} \\
G_{IJ} \rightarrow G_{IJ} - (G_{uI} G_{uJ} - B_{uI}B_{uJ})/G_{uu} \\
B_{IJ} \rightarrow B_{IJ} - (G_{uI}B_{Ju} - G_{uJ}B_{Iu})/G_{uu}
\quad , \quad 2 \phi \rightarrow 2 \phi -  \log G_{uu} \ .
\ea
\ee
On the other side the S-duality is implemented in the 4d theory. First we
have to go to the canonical frame by $G_{\mu\nu}^c=e^{-2\phi} G_{\mu\nu}$,
express the torsion by the axion $a(x)$ ($H =e^{4\phi}~^*da$) and
define a comlex scalar $S=a + i e^{-2\phi}$. Then the S-duality
says that the equations of motion  of (\ref{020}) are invariant
under\footnote{In general the $S$ duality is given by an $SL(2,{\bf R})$
transformation. We are considering here only the nontrivial
$SL(2,{\bf Z})$ part.}
\be040
\ba{l}
S \rightarrow -\frac{1}{S} \quad , \quad \\
F_{\mu\nu}^{(m)} \longrightarrow - Re S \,
F_{\mu\nu}^{(m)} \, - \, Im S \, (M L)_{mn} ~^*F_{\mu\nu}^{(n)}
\ea
\ee
where the definition of the matrices $M$ and $L$ can be found in the paper
of Sen\cite{se1} and the dual of the KK-gauge fields are
\be050 \
^*F_{\mu\nu}^{(m)} = \frac{1}{2} (\det G^c)^{-\frac{1}{2}} G^c_{\mu\mu'}
G^c_{\nu\nu'} \epsilon^{\mu'\nu'\lambda\rho} F_{\lambda\rho}^{(m)} \ .
\ee
Since the canonical metric remains unchanged under this transformation the
masses are the same too. An $S$-self-dual point is given by $|S|^2=1$, where
the axion changes only the sign and the dilaton remains unchanged.

Both $S$ \& $T$ duality are symmetries of one string model. In recent times
a complete different type of dualities is increasingly investigated. That is
the string/string duality in 6 dimension. This duality is a special case of
the string/p-brane duality which maps the string (1-dimensional objects) on
p-branes (p-dimensional objects) and acts in $p+5$ dimensional space. In 10
dimensions it is just the string/5-brane duality and in 6 dimensions the
string/string duality. There several aspects which make the string/string
duality attractiv to investigate. First it is an essential part in the
``unification'' of the five string theories (type I, IIA, IIB, both
heterotic models), see also Ref.\cite{sc}. Secondly, it interchanges classical
($\sim \alpha'$) and quantum corrections ($\sim e^{2 \phi}$). Therefore, if
one has a string model where the $\alpha'$ corrections are under controll
one knows the quantum corrections of the dual model. Implemented is this
duality by a field transformation in $D=6$
\be060
\ba{l}
H_{MNP} \rightarrow e^{-2 \phi} \ ^*H_{MNP} \\
G_{MN} \rightarrow e^{-2 \phi} G_{MN} \quad , \quad \phi \rightarrow -\phi \ .
\ea
\ee
All dualities are expected to be symmetries of the string theory. But
usually they are not respected by special backgrounds. Instead, they break
this symmetry. Nevertheless we can use these transformations to find new
dual backgrounds, which are equivalent from the string point of view. Some
special configurations, however, are explicitly invariant under
these dualities and these points in the field space correspond to
points of enlarged symmetries and often related to additional massless
modes\cite{hu/to}.

In this paper we are going to discuss the $T$ and $S$ duality for a special
background which allow self-dual points and includes many known black hole
and monopole solutions as special limits. We will find that in the self dual
limit our black holes or monopoles become massless. The model is
defined by the following 10d metric, antisymmetric tensor and
dilaton
\be070 \ba{l}
ds^2 =  2 F(x)\, du [ dv - \frac{1}{2} K(x) du +
\omega_I(x) dx^I ] - dx^I dx^I \\ \hat{B} = 2 F(x)\, du \wedge [ dv +
\omega_I (x) dx^I ] \qquad , \qquad e^{2 \hat{\phi}} = F(x) ~.
\ea
\ee
It can be shown\cite{ho/ts} that if $\partial^2 F^{-1}=
\partial^2K=0$, $\partial^j \partial_{[j} \omega_{i]}=0$ and $2
\hat{\phi}= \log F$ it is a solution of (\ref{010}) and do not receive
$\alpha'$ corrections in a proper renormalization scheme. Furthermore, this
model possesses unbroken supersymmtries and be embeded in $N=1$, $D=10$
supergravity\cite{ka}. It is the natural generalization of the
fundamental string ($K=\omega_I=0$) and the gravitational wave ($F=1,
\omega_I=0$) background. Since it is independent on $u$ and $v$ we can
dualize any non-null direction in the $(u,v)$ plane. A suitable form is that
we first shift $v$ by $v\rightarrow v + c u$ ($c=const.$), then dualize $u$
and afterwards revers the shift. Doing this and using (\ref{030}) we find
that the model does not change, only the harmonic functions $K$ and $F^{-1}$
are mixing
\be080
F^{-1} \rightarrow 2c -K \qquad , \qquad K \rightarrow 2c - F^{-1}
\ee
and the model is explicitly $T$-self-dual iff
\be090
K+F^{-1} = 2c \ .
\ee

\vspace*{5mm}

Now let us come to the 4d solution. We choose $v$ as time coordinate and
three of the transversal coordinates as spatial part,
i.e.\ $x^M=(v=t , x^i | x^r , u)$ and get\cite{be1}
\be100
\ba{l}
ds^2 =  e^{4\phi}(dt + \omega_i dx^i)^2 -  dx^i dx^i \quad , \quad
\partial_i a = \epsilon_{ijm} \partial_j \omega_m \\
\vec{A}^{(+)}_{\mu}
 = -\, \frac{1}{\sqrt{8}}  \,e^{-2 \sigma} ( \, 0 \, , \, 1 + KF \, )\,
   V_{\mu} \ , \
    \vec{A}^{(-)}_{\mu}
        =  -\, \frac{1}{\sqrt{8}}\, e^{-2 \sigma} ( \, 2 F \omega_r \, ,
	\, 1 - KF\, )\,
  V_{\mu} \\
e^{2\sigma} = K F - F^2 |\omega_r|^2 \quad , \quad
e^{-2\phi} = \sqrt{K F^{-1} - |\omega_r|^2}

\ea
\ee
with $V_{\mu} = F ( 1 , \omega_i)$, $a$ the axion and the vector field
$\vec{A}^{(+)}_{\mu}$ is sum of the KK gauge fields coming from the metric
and the antisymmetric tensor whereas $\vec{A}^{(-)}_{\mu}$ is just the
difference. So, the 4d solution has 7 gauge fields, one graviphoton
($\vec{A}^{(+)}_{\mu}$), 6 gauge fields belonging to the vector multiplet
($\vec{A}^{(-)}_{\mu}$) and 8 scalars correlated to the 8 harmonic functions
($K$, $F^{-1}$, $a$, $\omega_r$). These 8 harmonic functions determine
the solution completely. The simplest choice is
\be110
K = 1 + \frac{2 m}{r} \quad , \quad F^{-1} = 1 + \frac{2 \tilde{m}}{r}
\quad , \quad \omega_m = \frac{2 q^m}{r} \quad , \quad a = \frac{2 n}{r}
\ee
where $r^2 = x^2 + y^2 + z^2$. Inserting this we get for the metric
and dilaton
\be120
ds^2 = e^{4 \phi} \left( dt + 2 n \cos \theta d\phi \right)^2 -
\left( dr^2 + r^2 d\Omega^2 \right) \quad , \quad
e^{4 \phi} = \frac{1}{(1 + \frac{r_+}{r})(1 + \frac{r_-}{r})}
\ee
with $r_{\pm} = m + \tilde{m} \pm \sqrt{(m-\tilde{m})^2 + 4 |q_m|^2}$. This
is a Taub-NUT geometry, which is asymptotically not flat and has to have a
periodic time in order to avoid a whire singularity along the axes $\theta
=0, \pi$. All solutions are extremely charged solutions. Let us therefore
ignore this point in the further discussion.

The next possibility is that the harmonic functions are not singular in one
point ($r=0$) but along a ring. This can be done by choosing the real and
imaginary part of complex harmonic functions
\be130
\ba{ccc}
K = Re \left( 1 + \frac{ 2 m}{r_{\alpha}}\right)
& , &
F^{-1} =  Re \left( 1 + \frac{ 2 \tilde{m}}{r_{\alpha}}\right) \\
\omega^r = Re \left( \frac{ 2 q^r}{r_{\alpha}}\right)
& , &
a =  Im \left( \frac{ 2 n}{r_{\alpha}}\right) \ .
\ea
\ee
with $r_{\alpha}^2=x^2+y^2+(z-i\alpha)^2$.
These functions have a singularity in the plane $z=0$ along the circle
$x^2+y^2=\alpha^2$. Then the solution is a rotating black hole
\be140
\ba{l}
ds^2 =e^{4 \phi}
\left( dt + \frac{2 n \alpha \sin^2 \theta}{R} d\phi \right)^2 -
d\vec{x}^2 \  , \
e^{-4 \phi} =  (1 + \frac{r_+}{R})(1 + \frac{r_-}{R})
\  , \  R = \frac{r^2 + \alpha^2 \cos^2 \theta}{r}
\ea
\ee
This solution is asymptotically flat and we can define the charges and
get for the mass, angular momentum and charges
\be150
\ba{l}
M= \frac{1}{4} (r_+ + r_-) = \frac{1}{2} (m + \tilde{m}) \quad  , \quad
J= n\, \alpha \\
\vec{Q}^{(+)}=\sqrt{2} ( 0 , M ) \quad , \quad
\vec{Q}^{(-)}= \frac{1}{\sqrt{2}} ( -2 q^r , (\tilde{m}-m)) \ .
\ea
\ee
As expected, we see that the right-handed part (graviphoton) saturates the
Bogomol'nyi bound.

\vspace{3mm}

Now we have the 4d solutions and can look on the self-dual cases. From
(\ref{090}) and (\ref{130}) it follows that at the $T$-self-dual point
(with $c=1$) $m+\tilde{m}=0$ and therefore $M=0$, i.e.\ the black hole
becomes massless and therefore the right-handed sector is uncharged
($Q^{(+)}=0$).  Note that the $T$ duality here corresponds {\bf not}
to an isometry direction in the internal space. Instead, it is a
mixing between the time ($v=t$) and the internal coordinate $u$. Thus
it is different from the $T$ duality correlated to $O(6,6)$ moduli
space. Furthermore, we see that in this case we have an additional
singularity at $r = -r_-$. This singularity is a consequence of the
dimensional reduction and exhibits a gravitational
repulsion\cite{li/ka}. The 10d theory is non-singular at this
point. In addition, we have still the naked singularity at $R=0$.  But
this singularity cannot be removed by uplifting to higher
dimensions. However, by assuming that the fields depend on more than
three coordinates and making them periodic the naked singularity
disappears\cite{ho/se}.

We can also identify this configuration as state in the string spectrum. As
usual we consider the mass formular
\be160
M^2 = \frac{1}{2} \left( \, |\vec{Q}_e^{(+)}|^2 + 2(N_R -\frac{1}{2})\,
\right) = \frac{1}{2} \left(\, |\vec{Q}_e^{(-)}|^2 + 2(N_L - 1 )\, \right)
\ee
where $N_{R/L}$ are the oscillator modes. For our self-dual configuration we
have $M= Q^{(+)}=0$ and therefore $N_R=\frac{1}{2}$. Since $Q^{(-)}\neq 0$
the only possibility for the left moving sector is given by $N_L = 0$. This
elementary state is electrically charged. The magnetic
counterpart\cite{be/ka} appear then as soliton excitation related to the
electric solution.

Next, we look on the $S$-self-dual point, which was given by
\mbox{$|S|^2=a^2+e^{-4\phi}=1$}. Looking on the harmonic functions, we
see that this case is only possible for the Taub-Nut geometry. We obtain
for the parameter the restrictions
\be170
m+\tilde{m}=0 \quad , \quad n^2 = \tilde{m}^2 + |q|^2
\ee
and thus it coincides with the $T$-self-dual point with an additional
restriction for the parameter $n$.

Finally let me add an remark concerning the string/string dual of this
solution. We know\cite{du} that the scalar fields are interchanging, e.g.\
the dilaton and modulus field. The KK gauge fields from metric remains
unchanged and the KK gauge field from the antisymmetric tensor besomes
magnetic. But the crucial point is that the canonical 4d metric
remains unchanged and thus the dual masses are the same, which means
that this massless state is maped again on a massless state on the
type IIA side. We will come back in more detail about this
solution in a forthcoming paper\cite{be/do}.

\section{Acknowledgements}
The author would like to thank the Physics Department of the
Stanford University for the hospitality where parts of this work were
done. This work is supported by the DFG and a grant of the DAAD.

\end{document}